\documentstyle[aps,epsfig,tighten]{revtex}
\def\Journal#1#2#3#4{{#1} {\bf #2}, #3 (#4)}

\def\AP{{\em Ann. Phys.}}

\def\EPJC{{\em Eur. Phys. J.} C}

\def\NPB{{\em Nucl. Phys.} B}

\def\PLB{{\em Phys. Lett.} B}
\def\PRL{\em Phys. Rev. Lett.}

\def\PRD{{\em Phys. Rev.} D}

\def\PREPC{{\em Phys. Rep.} C}

\def\RMP{{\em Rev. Mod. Phys.}}
\def\RNC{{\em Riv. Nuovo Cimento}}

\def\YA{{\em Yad. Fiz.}}

\newcommand{\be}{\begin{equation}}
\newcommand{\ee}{\end{equation}}
\newcommand{\bea}{\begin{eqnarray}}
\newcommand{\eea}{\end{eqnarray}}

\newcommand{\hf} {{1\over2}}
\newcommand{\nonu}{\nonumber\\}

\def\eq#1{(\ref{#1})}
\def\ra{\rangle}
\def\la{\langle}

\begin{document}
\title{Confinement as crossover}

\author{Janos Polonyi\thanks{polonyi@fresnel.u-strasbg.fr}}
\address{Laboratory of Theoretical Physics, Louis Pasteur
University\\
3 rue de l'Universit\'e 67084 Strasbourg, Cedex, France}
\address{Department of Atomic Physics, L. E\"otv\"os University\\
P\'azm\'any P. S\'et\'any 1/A 1117 Budapest, Hungary}
\date{\today}
\maketitle
\begin{abstract}
The order parameter of confinement together with
the haaron model of the QCD vacuum is reviewed and it is
pointed out that the confining forces are generated by the
non-renormalizable, invariant Haar-measure vertices of the
path integral. A hybrid model is proposed for the description of the
crossover leading to the confining vacuum. This scenario suggests
that the differences between the low and the high temperature
phases of QCD should be looked for in the quark channels instead of the
hadronic sector.
\end{abstract}
\section{Introduction}
Having no systematic derivation of the confining
forces in pure Yang-Mills theories the studies of the long
range properties of the hadronic vacuum are usually based on
model computations \cite{bag}-\cite{haaron}.
In order to get closer
to the understanding of the problem it is obviously better to
use models which contain at least partially the original
gluonic degrees of freedom.

The bag model \cite{bag} is based on
weakly interacting quarks and gluons, the confinement
is realized by the difference of the energy density of
two different vacua: inside and outside of the bag. No colored
degrees of freedom are supposed to exist outside.
The boundary of the bag, considered as a dynamical degree
of freedom is obviously non-renormalizable. In the Abelian dual
superconductor model \cite{dual} the Schwinger relation which asserts
that the electric and the magnetic charges are inversely proportional
to each other indicates that the magnetic condensate is due to a
non-renormalizable coupling constant. The stochastic confining
model \cite{stoch} does not shed light on this question since it is based
on the cluster expansion leaving the origin of the correlations
an open question.
The invariant Haar-measure terms of the functional
integral which yield the string tension within the
haaron model \cite{haaron} are non-renormalizable, as well.

It is worthwhile noting that haaron model is the only one
where the non-renormalizable term which is responsible for the
string tension is already present in the original asymptotically free
Yang-Mills Lagrangian. In fact, the gauge invariant, non-perturbative
lattice regularization is based on the invariant Haar measure for
the gauge group valued link variables.
The logarithm of the Haar measure, treated as a
local interaction potential in the action is non-polynomial
and thereby non-renormalizable. Nevertheless the cut-off
can be removed by suppressing these vertices
sufficiently fast in the continuum limit \cite{riesz}.

We are confronted with an interesting possibility:
How can it happen that the leading infrared force of the Yang-Mills
theory comes from non-renormalizable vertices? The string tension,
being a dimensionful parameter, can be generated by a relevant
operator only. Since the non-renormalizable terms are irrelevant
these vertices influence the interaction at the cut-off scale
only and could have been left out from the theory according to the
universality.
The solution of this apparent paradox is rather simple \cite{cren}:
Any theory with internal scale has at least two scaling regimes,
an UV and an IR one, separated by a crossover at the internal scale.
The non-renormalizable vertices are indeed irrelevant in the
UV scaling regime but they might become relevant at the
IR side of the crossover, in the IR scaling regime, where
the IR forces are generated. 

Section II overviews briefly the symmetry and the order parameter
related to the confinement. An effective theory, the haaron model,
to describe confinement as a destructive interference is mentioned
in Section III. The lesson of the manner the confining force
is generated in this model is discussed in Section IV.
The finite temperature aspects of confinement as a crossover phenomenon
are touched upon in Section V. Finally, Section VI is for the
conclusion.

\section{Order parameter and destructive interfence}
The order parameter for confinement is given in terms
of the analytically continued massive quark propagator,
\be\label{orderp}
\Omega(x,t)=tr\la\psi(x,t)\bar\psi(x,t+i\beta)\ra,
\ee
where the trace is over the color and the spin indices and $\beta=1/T$. 
Note that for infinitely heavy quarks in a time independent environment
this expression reduces to the Polyakov line
\be
\omega(x)=trPe^{i\int_0^\beta d\tau A_0(x,\tau)}
\ee
up to a constant multiplicative factor, where $A_0(x,\tau)$ is the
temporal component of the Euclidean gauge field.

$\Omega$ displays the status of the symmetry with respect to the
center of the gauge group which is the group $Z_n$ for the gauge group
$SU(n)$. The order parameter can be easily introduced even for
Yang-Mills models in continuous Minkowski space-time \cite{mech}
and it remains a manifestly gauge invariant, well defined observable.

The dynamical quarks make the picture more complicated and we should
distinguish two competing confinement mechanisms, a hard and a soft
one. Both are driven by the increase of the effective coupling strength
as the color charges are separated.
The hard confining mechanism of the Yang-Mills models is responsible
for the flux tube formation and the linearly rising potential between
a static quark-anti quark pair. The soft mechanism is due to the
Dirac-sea polarization and, similarly to the supercritical vacuum
of QED \cite{supcrit}, shields the isolated quarks \cite{gribov}.
The soft mechanism cuts short the hard one and saturates
the linearly rising potential when the flux tube
between a static quark-anti quark pair is broken by the polarization
of the Dirac-sea.

This "deconfining " vacuum-polarization effect appears in the
dynamics of our order parameter, as well: The formal center
symmetry is broken by the fermion determinant in the
grand canonical ensemble. But it is easy to see that the Legendre
transformation of the baryon number between the canonical and
the grand canonical ensemble is ill defined in the thermodynamical
limit due to the confinement mechanism. In fact, the free energy
is infinitely large for states with non-vanishing triality\footnote{
The $n$-ality of a multi-quark state with $N$ quarks and $\bar N$
anti-quarks in an $SU(n)$ gauge model is defined as $t=N-\bar N({\rm mod} n)$.}
which turns the free energy into a non-differentiable function of
the baryon number density in the thermodynamical limit,
and the control of the baryon number by a chemical
potential into a highly non-trivial problem \cite{canonical}. It is the
canonical ensemble for the triality \cite{canonical}, \cite{canonicao}
which should rather be used
in this case and this ensemble is formally center symmetrical.
But this formal symmetry is broken spontaneously at low temperatures
\cite{canonical}. At high temperature the center symmetry is broken
dynamically by the gluon kinetic energy. There is
no reason to expect that the two unrelated symmetry breaking mechanisms
would generate the same expectation value for $\Omega$ thereby
\eq{orderp} remains to be an order parameter which
experiences a non-analytic dependence on the environmental variables
at the deconfinement transition \cite{canonical}.

The dynamical picture of confinement with $SU(n)$ as color gauge group
is the following \cite{mech}: The configuration space for global gauge
rotations, $SU(n)/Z_n$, is multiply connected. Its fundamental group,
the center $Z_n$ can be used to lump the time-dependent gluon field
configurations into $n$-tuples in such a manner that the
trajectories of an $n$-tuple correspond to the same initial or end
points in the multiply connected space $SU(n)/Z_n$ but differ on the
covering space, $SU(n)$. The center symmetry of the pure gluonic system
makes the action $S$ of the trajectories of an $n$-tuple degenerate
in the absence of quarks. Thus the contribution of $n$ trajectories
of an $n$-tuple to the transition amplitude is
\be
{\cal A}_E=ne^{-S},~~~{\cal A}_M=ne^{iS}
\ee
in Euclidean and Minkowski space-time.
When a spectator quark is propagating along with the gluons
then it picks up the $Z_n$ phase of the center transformation
and the contribution of an $n$-tuple is vanishing
due to the destructive interference between the homotopy classes,
\be\label{interf}
{\cal A}_E=\sum_{\ell=1}^ne^{i{2\pi\over n}\ell-S},~~~
{\cal A}_M=\sum_{\ell=1}^ne^{i{2\pi\over n}\ell+iS}.
\ee
Notice that the non-positive definite phase factor comes
from the projection operator which is supposed to install Gauss' law
and may lead to a destructive interference even for imaginary time.
The semiclassical expansion, saturated by Wu-Yang monopoles
in the Prasad-Sommerfeld limit \cite{perv} supports the confinement as a
destructive interference phenomenon, as well.

For an $SU(2)$ gauge model the center symmetry expresses the
invariance of a three vector under rotation by angle $2\pi$
and the destructive interference is due to the factor $-1$
the spinors of the fundamental representation collect during
a rotation by $2\pi$.

The high temperature deconfining transition is due to the too
high kinetic energy barrier for gluons
to follow the trajectories in the whole homotopy class \cite{mech}. In the
high temperature phase "there is no time" to realize all homotopy
class, the destructive interference is prohibited and quarks can
propagate. Note that the kinetic
energy driven dynamical symmetry breaking occurs at high energy
or in short time processes contrary to the potential energy
governed spontaneous symmetry breaking which is observed at low energy
or long time. When dynamical quarks are present then the
spontaneous breakdown of the center symmetry selects a
homotopy class which dominates the sum \eq{interf} and leads
to the screening of the isolated triality charge.

\section{Haaron model}
We start this brief summary of the haaron model with a remark about
the importance of keeping the exact gauge invariance in a computation
to extract the string tension in Yang-Mills theory. Suppose
that gauge invariance is implemented in an approximate manner and
the state with a static quark-anti quark pair is
$|q\bar q\ra_0+|q\bar q\ra_1$ where $|q\bar q\ra_0$ has the proper
transformation rule under gauge transformations and $|q\bar q\ra_1$
not. The non-covariant component $|q\bar q\ra_1$ appears
to contain uncontrollable charge distribution. The expression
\be
\left(\la q\bar q|_0+\la q\bar q|_1\right)H
\left(|q\bar q\ra_0+|q\bar q\ra_1\right)
=\la q\bar q|_0H|q\bar q\ra_0+\la q\bar q|_1H|q\bar q\ra_1
\ee
for the static potential shows that the charges in the component $|q\bar q\ra_1$
will break the flux tube for a sufficiently large separation
of the test charges and saturate the potential. The gauge non-covariant
components shield off the string tension when they are present with any
small amplitude.\footnote{This does not present serious
problem in QED where the gauge non-covariant contributions are
not gaining importance by non-perturbative effects in the absence
of the confining forces.}

There is another indication of the strong relation between gauge
invariance and the confining forces. The effective theory for the
Polyakov line obtained in the strong coupling expansion
shows that the minimum of the effective potential
is at a vanishing value of the order parameter at low temperature
due to the presence of the invariant Haar-measure for the gauge field
\cite{effth}. The replacement of the Haar-measure
with a non-compact measure may keep the gauge invariance intact at any
finite order of the loop expansion but certainly would break it
at a non-perturbative level.

Guided by these remarks we wish to base our effective theory
for the confining forces on the invariant Haar-measure
in the path integral. Consider $SU(2)$ Yang-Mills theory for
simplicity where the center transformation amounts to
$\Omega\to-\Omega$. The lattice regularized path integral
is given as
\be
\int D_H[aA_\mu(x)]e^{-S_{YM}[aA_\mu(x)]},
\ee
where $a$ stands for the lattice spacing. The invariant integration
measure for $A_0^j(x)=u(x)\omega^j(x)$, $j=1,2,3$, $(\omega^j(x))^2=1$
can be written as
\be\label{hmeas}
D_H[aA_0(x)]=D[\omega(x)]D[u(x)]e^{\sum_x\log\sin^2au(x)}
\ee
in terms of the flat integration measure $D[u(x)]$ for $u(x)$.
The center transformation, $u(x)\to u(x)+\pi/a$, performed
at a given equal-time hypersurface is a symmetry of the periodic
potential $a^{-4}\log\sin^2au$ appearing\footnote{The factor
$a^{-4}$ is to compensate the space-time integration volume $d^4x$
of the potential in the action. It makes the measure term
vanishing in dimensional regularization.} in \eq{hmeas}.
Let us integrate out the UV modes from the path integral and lower the
cut-off. This step makes the dimensional transmutation explicit and
induces an effective
gauge theory model with dimensional parameters. First we choose a
gauge in this theory where the temporal component $A_0(x)$ is diagonal,
$\omega^a(x)=\delta^{a,3}$ and then we set the non-diagonal
components of the gauge field to zero,
leaving behind a compact $U(1)$ gauge model. The Feynman gauge is
chosen for this model. As the final step, we set the spatial
component of the Abelian gauge field to zero. What we find at the end
is an effective theory for the diagonal, temporal component of the
original gauge field, $u(x)$. The corresponding effective
action will be approximated in the framework of the gradient
expansion by the form
\be
S_{eff}[u]=\int d^4x\left[\hf(\partial_\mu u(x))^2-V(u(x))\right].
\ee
The potential, the remnant of the invariant Haar-measure
is periodic $V(u+2\pi/\ell)=V(u)$. The periodicity reflects
the discrete center symmetry of the original theory and allows
the Fourier representation $V(u)=\sum_nv_m\cos m\ell u$.
The center symmetry requires
that the symmetry $u\to u+2\pi/\ell$ is respected by the vacuum.
The obvious consequence of this symmetry is the absence of any barrier
in the effective potential between the periodic minima. This leads
to the flattening of the effective potential for $u$ and the masslessness
of the field $u(x)$.

It is well known that the sine-Gordon model is equivalent with a
Coulomb gas. The simplest way to see this is to expand the generating
functional in the coupling constant $v$,
\bea\label{resum}
Z[\rho]&=&\int D[u]e^{-\hf u\cdot G^{-1}\cdot u+iu\cdot\rho+\hf v_m\int dx
(e^{im\ell u(x)}+e^{-im\ell u(x)})}\nonu
&=&\sum_n{(v_m/2)^n\over n!}\prod_{j=1}^n\int dx_j\sum_{\sigma_j=\pm1}
\int D[u]e^{-\hf u\cdot G^{-1}\cdot u+iu\cdot(\rho+\sigma)}\nonu
&=&e^{-\hf\rho\cdot G\cdot\rho}\sum_n{(v_m/2)^n\over n!}
\prod_{j=1}^n\int dx_j\sum_{\sigma_j=\pm1}
e^{-\hf\sigma\cdot G\cdot\sigma}e^{-\rho\cdot G\sigma},
\eea
where $G$ is the massless propagator and
$\sigma(x)=m\ell\sum_{j=1}^n\sigma_j\delta(x-x_j)$.
This is the grand canonical partition function of a four dimensional
gas of particles interacting with the inverse of the massless propagator,
the Coulomb potential. The first exponential in the last line represents
the perturbative self-interaction of the external source $\rho$,
the second stands for the self-interaction of the particles and
finally the third one describes the interaction between the source
and the particles. Let us ignore the inter-particle forces and the
partition function for non-interacting particles can be resummed,
\be
Z[\rho]\approx e^{\hf\rho\cdot G\cdot\rho+\hf v_m\int dx
\cos(im\ell\int dyG(x-y)\rho(y))}.
\ee
These steps, repeated for each Fourier modes give
\be\label{freeh}
Z[\rho]=\int D[u]e^{-\int d^4x[\hf(\partial_\mu u(x))^2-V(u(x))
-i\rho(x)u(x)]}\approx
e^{-\hf\rho\cdot G\cdot\rho+\int dxV(i\int dyG(x-y)\rho(y))}.
\ee
The particles representing the vertices of the perturbation
expansion are called haarons since their contributions
come from the invariant measure of the original path integral.

It is worth mentioning three applications of the partial resummation
\eq{freeh}. The leading long range part of the static potential
between a quark-anti quark pair separated by $x$ is
\be
-\int d^3yV\left({i\over|y|}-{i\over|y-x|}\right)\approx-2V''(0)|x|,
\ee
giving the string tension
\be\label{strt}
\sigma=-2V''(0).
\ee
Notice that in the
center symmetry broken phase there is no protection against mass generation
and the massive propagator does not give linearly rising potential.
Thus the measure term which gives vanishing contribution in the
UV regime and is included to assure the full gauge invariance only
actually generates the leading long range force. Since it generates
a new dimensional parameter, the string tension, the measure
term must be relevant in the IR regime.

The second application of the resummation gives a confining version
of the NJL model. We start with the Lagrangian
\be
L=\hf(\partial_\mu u)^2-V(u)+i\bar\psi\partial_\mu\gamma^\mu\psi-igj~,
\ee
where $j=u\bar\psi\gamma^0\sigma_z\psi$ and perform the free haaron gas
resummation yielding
\bea
L_{eff}&=&i\bar\psi(x)\partial_\mu\gamma^\mu\psi(x)
-V\left(ig\int d^4yG(x-y)j(y)\right)\nonu
&\approx&i\bar\psi(x)\partial_\mu\gamma^\mu\psi(x)
-\hf g^2V''(0)\int d^4yj(x)G_2(x-y)j(y)\nonu
&=&i\bar\psi(x)\partial_\mu\gamma^\mu\psi(x)+g\sqrt{-V''(0)}j(x)\phi(x)
+\hf\phi(x)\Box^2\phi(x)~,
\eea
where
\be
G_2(x-y)=\int d^4zG(x-z)G(z-y)=\int{d^4p\over(2\pi)^4}
{1\over p^4}e^{-ip(x-y)}~,
\ee
and the auxiliary field $\phi(x)$ was introduced in order to
render the Lagrangian local. Notice that the $1/p^4$ propagator
of the auxiliary field, coupled to the quarks in the same manner as $u(x)$,
confines the color charges with a linearly rising potential and the
string tension is \eq{strt}.

The third application of the resummation is the computation
of the quenched quark propagator. The grand canonical partition function,
the last line of Eq. \eq{resum} when $\rho$ is replaced by the quark current
$j(x)$ shows that the quarks are propagating in the imaginary long range field 
\be
u_{y,n}(x)={in\ell\over4\pi^2(x-y)^2}
\ee
of the haarons. After performing the
Wick rotation into Minkowski space-time this external field becomes
real. The destructive interference between the homotopy classes
appears in this effective model as the destructive interference
between the scattering processes of a quark off the gas of haarons.
The long range haaron field makes the phase shift diverging and
the fast rotating phase of the scattered state
cancels the quark propagator when the averaging over the haaron
distributions is performed. In order to understand the propagation
of a meson qualitatively let us assume that the haaron field
at $x$ and $y$ is identical or completely uncorrelated when $|x-y|<\xi$
or $|x-y|>\xi$, respectively where $\xi\approx\Lambda^{-1}_{QCD}$
is the correlation length
of the haaron gas. As long as the quark and the anti-quark of the meson
propagates within the distance $\xi$ the phase shift suffered by them
is canceled and the haarons do not influence much the propagation.
When the color charges are separate from each other more than
$\xi$ then the statistically independent phase shift suppresses
the amplitude. The result is that the world lines can not separate
more then the distance $\xi$, the confinement radius.

\section{Crossover in the vacuum}
Let us consider the thought-experiment when the hadronic matter
is viewed by a microscope of adjustable space-resolution. When
details below the distance scale $\Lambda^{-1}_{QCD}$ are considered
we find partons, i.e. quarks and gluons. As the resolution becomes
worse and details on the scale well above $\Lambda^{-1}_{QCD}$ are
seen only then hadrons and glueballs are found. The interactions
between quarks and gluons on the one hand, and between hadrons on the
other hand, are very different. This difference can simply be
recorded by following the scale dependence generated by them.

There are at least two different scaling regimes in any non-scale
invariant theory, an UV and an IR one separated by a crossover
at the internal scale of the theory, $\Lambda^{-1}_{QCD}$ in our case,
c.f. Fig. \ref{ev}. The
UV scaling reflects asymptotically free forces between quarks and
gluons in the UV regime and short ranged Yukawa interactions
among the asymptotic states, hadrons, on the IR side. In pure
Yang-Mills theory glueballs are the asymptotic states in the IR
and color charges remain strongly bound by the linearly rising
potential. What was surprising in the haaron model picture is
that the measure vertices of the action which are non-renormalizable,
i.e. irrelevant in UV scaling regime can generate the leading
long range force. There must be a change in the behavior of the
measure vertices as we move towards the IR directions which explains
their increased importance in the confining forces. The most
natural scenario is that these operators, being irrelevant in the
UV scaling regime become relevant in the IR side of the crossover.

This scenario raises a more general question, the possibility that
non-renormalizable operators might play an important role in low
energy physics. It is easy to see that this surprising phenomenon
does not take place in models with mass gap $m\not=0$. These models
display a correlation length $\xi\approx1/m$ and the evolution of the
running coupling constant slows down at distance $x \gg \xi$.
In fact, the evolution of the coupling constants is driven by the
contribution of the modes around the running cut-off and the
fluctuations at the scale $x \gg \xi$ are suppressed
by $\exp(-x/\xi)$. The absence of runaway trajectories of the
renormalization group flow indicates that all non-Gaussian operators
are irrelevant in the IR scaling regime\footnote{An irrelevant coupling
constant may naturally be important if its fixed point value is
not small.}.

Theories without mass gap may develop new relevant operators
by the help of collinear or simple IR divergences which may drive
the run-away trajectories. The $\phi^4$ model in the mixed
phase possesses a non-renormalizable operator which is relevant
at low energies \cite{grg}. The condensation mechanism
in general can easily generate radically new scaling laws
\cite{tree}. When gauge symmetry is protecting against mass generation
then the four fermion interaction, the effective vertex
responsible for the emergence of the BCS phase, turns out to be relevant
at low energies \cite{shankar}. The interaction vertices between
hadronic states are irrelevant in QCD because the colorless sector
is massive. The lesson of the haaron model is that the integral measure
vertices become relevant in the IR scaling regime of the colored
channels.

There are two ways to deal with changing scaling laws. The
phenomenological approach is the matching where one introduces different
models for the UV and the IR scaling regimes and tries to match them
at the crossover. A more instructive procedure can be constructed by recalling
one of the rules of the renormalization group studies: All coupling
constants which are generated by the blocking and might turn out
to be important should be present in the action from the very
beginning. In fact, the renormalization group flow is a reliable
source of information about the interactions only if the
truncation of the space of Hamiltonians does not remove
important pieces. This principle, applied to QCD suggests the
introduction of colorless composite operators which control
the hadronic states,
\be
Z=\int D[\bar\psi]D[\psi]D_H[A_\mu]D[\bar\chi]D[\chi]D[\bar\Psi]D[\Psi]
e^{iS_{QCD}[\bar\psi,\psi,A_\mu]+iS_H[\bar\chi,\chi,\bar\Psi,\Psi]
+i\int dx(G_\chi\bar\chi\psi\psi\psi+G_\Psi\bar\Psi\bar\psi\psi+h.c.)},
\ee
the fields $\chi$ and $\Psi$ correspond to baryon and meson
states and $S_H[\bar\chi,\chi,\bar\Psi,\Psi]$ is the action for
a hadronic field theory.
Since we are interested in the low energy phenomena we
can fix the original cut-off at a sufficiently high but finite energy
scale $\Lambda_0$. The coupling constants $G_\chi$ and $G_\Psi$
govern the strength of interactions between the
hadronic and the colored states and their initial value is
$G_\chi(\Lambda_0)=G_\Psi(\Lambda_0)=0$, together with the hadronic coupling constants in
$S_H$. This scheme is not a double counting since it is casted
in the path integral formalism, it is a
possible parameterization of the effective action.

Such a hybrid model should hold the key to the understanding of
the confinement phenomenon because it offers a singularity-free
description of the crossover. As the cutoff is lowered the
non-renormalizable coupling strengths remain small and the
asymptotically free coupling $g$ grows. When the crossover is reached
then $g$ explodes in perturbative QCD and an IR Landau-pole arises because
the long range correlations of the ground state are supposed to be
generated by the asymptotically free vertices. But such a
hybrid model offers the following alternative. In the presence
of operators which are important in the IR scaling regime
there is a chance that $g$ stays finite because the desired
long range correlations can be established first in the colored
and after that in the neutral sector by the renormalization of the measure term
as in the haaron model and the hadronic coupling constants,
respectively.

\section{Crossover at high temperature}
\begin{figure}
\begin{minipage}{5cm}
\epsfxsize=10cm
\epsfysize=5cm
\centerline{\epsfbox{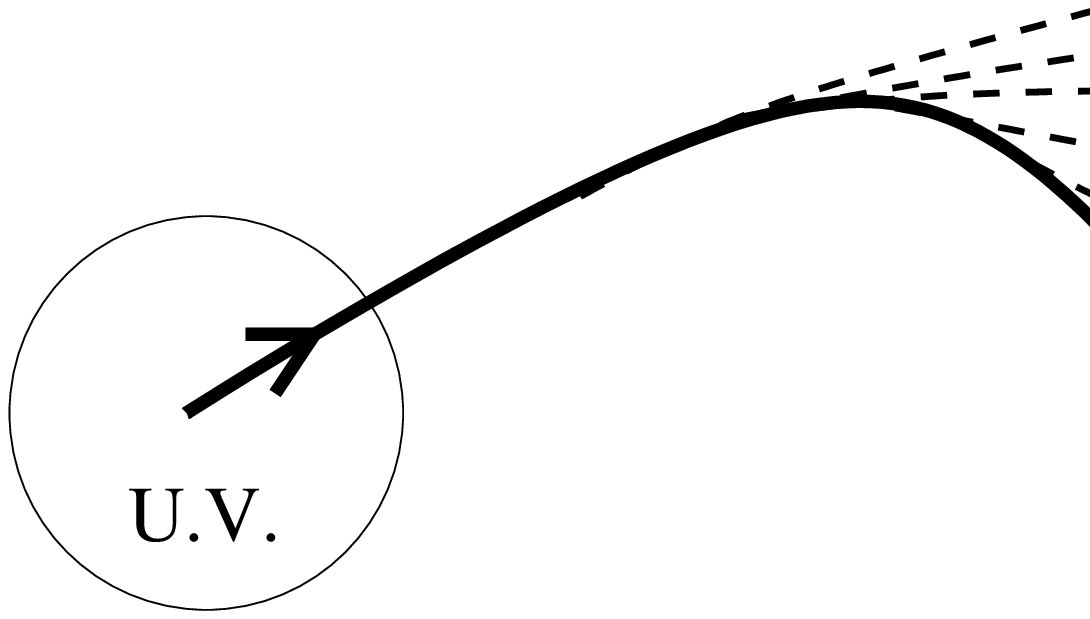}}
\end{minipage}
\caption{\label{ev}The renormalization group flow of QCD shown
for different temperatures. The solid line starting at the
UV and ending at the IR fixed point, encircled by the limit of
asymptotic scaling regimes, corresponds to $T=0$. The crossover
separating the UV and IR scaling regimes is at $Cr$,
$x\approx\Lambda^{-1}_{QCD}$. The flow at finite temperature
displays temperature dependence when $x>1/T$ and gives
two qualitatively different IR fixed point manifolds
for $T<T_{dec}$ and $T>T_{dec}$.}
\end{figure}

The real RHIC experiment is different, we are interested
in the long distance correlations and quasiparticle structure
at high temperature assuming that thermal equilibrium is
an acceptable approximation. How does the temperature
modify the scaling laws and the renormalized trajectory?
It is obvious that the renormalized trajectory is
in good approximation temperature independent at the
observational length scale $x \ll 1/T$ and the temperature
induced effects show up around $x\approx1/T$, as shown qualitatively in
Fig. \ref{ev}. For $T<T_{dec}$ the clusterization of the
color charges can be best understood as the impossibility
of screening the $1/3$ color charge of a quark by
multi-gluon states whose color charge is sum of integers,
$\sum\pm1\not=1/3$. The absence of screening mechanism
leads to confining forces. The deconfining phase transition
can be characterized in the Hamiltonian description
by the improper implementation of the Gauss' law
projection operator which does not exclude certain states with
infinitely many gluons. These states carry the color charge
of a quark or anti-quark \cite{mech}. The result is
the possibility of screening a quark color charge by a gluon cloud
whose wave functional is multi-valued,
the rearrangement of the infinite sum $\sum\pm1$ in such an
order that it converges to $1/3$. The effect of the temperature at the
deconfining transition is the removal of the linear
potential between triality charges by vacuum-polarization,
a mechanism similar to the polarization of the Dirac-sea
when dynamical quarks are present. A sort of soft confinement
mechanism is operating in the high temperature phase of the
pure glue system. The deconfined quarks are rendered colorless
and only their flavor quantum numbers reveal their quark content.

The renormalization group flow of the pure glue system should have
two different manifolds of IR fixed points, for $T>T_{dec}$
and $T<T_{dec}$, as shown in Fig. \ref{ev}. The high temperature
fixed points should be qualitatively similar to those of full
QCD at low temperature, the role of quarks are being played
by gluon states with multi-valued wave functionals.

For full QCD the difference between the low and the high
temperature fixed point manifolds is more subtle since the
soft confinement mechanism is operating in both phases, by
means of quark or gluon states with multi-valued wave functionals.
I believe that the quasiparticles of the high temperature fixed point
are similar to those of the low temperature phase except that
a new quark "flavor" appears in the form of gluonic states
with multi-valued wave functionals. The difference between the
real and this fake quark can be detected by electro-weak
currents only, the color charges being equivalent.
The main features of this scenario are the screening of the
color charge of the deconfined quark and the presence of the
usual hadronic bound states. As of the former, a careful
numerical study should be carried out measuring the gluonic
color charge polarization around a deconfined quark.
There are two indirect numerical evidences supporting
the latter conjecture, the presence of the usual hadronic
bound states. The first is the observation that the
spacelike string tension is to a large extent temperature independent
and remains non-vanishing even at high temperature \cite{spacel},
indicating that the equal time long-range correlations of the
multi-quark states do not change at the deconfinement phase transition.
Another result, indicating the unimportance of the string tension from the
point of view of the structure of the hadronic states at $T=0$ is that
the hadronic structure functions are qualitatively reproduced
after cooling, a modification of the gluonic configurations
which removes the string tension.

\section{Conclusion}
It was argued in the framework of the haaron model that the leading
long range forces between a quark-anti quark pair are generated by
non-renormalizable vertices. This phenomenon motivates a look into
the confinement problem following the strategy of the renormalization
group and suggests that the confinement characterizes the IR scaling
regime.

A lesson learned in dealing with the renormalization group is that
all important operators should be present in the initial
Hamiltonian even with vanishing coupling strength in order to
understand the appearance of the dynamically generated, new
kind of forces. Such a point of view motivates a hybrid
model which contains both the quark-gluon and the hadronic fields.
Their difference is set by the initial condition
for the renormalization group flow only: a finite value for the
asymptotically free coupling constant and a vanishing strength for the
hadrons. The non-renormalizable measure term is
supposed to generate a crossover in this model where the
hadronic coupling constants turn on and induce the interactions
of nuclear physics.

Such a description of the long range structure, together with the
screening mechanism of the quark color charges by gluons
available at high temperature suggests that the main
difference between the high and the low temperature phases
is not in the hadronic but rather in the quark sector of QCD.
Possible examples are the following: The chromomagnetic monopoles,
being "hedgehog" configurations relate color and spin. The gluon
polarization cloud around a deconfined quark with even or odd number
of monopoles possesses integer or half-integer spin, respectively. In
this manner the violation of the superselection rule for the
charge induces a similar violation for the spin and the
deconfined quark state is actually the sum of components with
Bose and Fermi statistics \cite{mech}. Another triality-related
effect is the deviation of the temperature of the quark and gluonic
degrees of freedom at the deconfinement phase transition point,
$T_q=T_{gl}/3$ \cite{thqua}.

\end{document}